%% file: main.tex
\documentclass[3p,times,twocolumn]{elsarticle}

\usepackage{color}
\usepackage{caption}
\usepackage{subcaption}

\usepackage{longtable}

\usepackage{amssymb}
\usepackage{amsthm}

\input{./src/usonics2023style-inc}

\begin{document}

\input{./src/usonics2023style-005-frontmatter}

\input{./src/usonics2023style-010-intro}
\input{./src/usonics2023style-210-da}
\input{./src/usonics2023style-020-related}
\input{./src/usonics2023style-220-luvt}
\input{./src/usonics2023style-270-alumspe}
\input{./src/usonics2023style-230-example}
\input{./src/usonics2023style-030-dataset}
\input{./src/usonics2023style-240-artificial}
\input{./src/usonics2023style-250-st}
\input{./src/usonics2023style-260-prop}

\input{./src/usonics2023style-310-table-cygan}

\input{./src/usonics2023style-040-meth}

\input{./src/usonics2023style-050-exp-cygan}

\input{./src/usonics2023style-320-table-modeldetatil}
\input{./src/usonics2023style-330-table-dataset}
\input{./src/usonics2023style-340-table-pred_result1}
\input{./src/usonics2023style-350-table-pred_result2}
\input{./src/usonics2023style-360-table-gradcam1}
\input{./src/usonics2023style-060-exp-imgcls}

\input{./src/usonics2023style-070-concl}

\section*{Declaration of Competing Interest}
The authors declare that they have no known competing financial
interests or personal relationships that could have appeared to influence
the work reported in this paper.

\section*{Acknowledgment}
This work was supported by SECOM Science and Technology Foundation and JSPS KAKENHI (C)(21K0423100).
This work used computational resources provided by Kyoto University and
Hokkaido University through Joint
Usage/Research Center for Interdisciplinary Large-scale Information
Infrastructures and High Performance Computing Infrastructure in Japan
(Project ID: jh220033).




\bibliographystyle{elsarticle-num}
\bibliography{usonics2023style-bib,kato2205,nakajmiya-bib}



\end{document}

%% file: src/usonics2023style-inc.tex
\usepackage{amsmath}
\usepackage{amssymb}
\usepackage{amsthm}
\usepackage{bm}
\usepackage{boxedminipage}
\usepackage{dsfont}
\usepackage{graphicx}
\usepackage{lineno}
\usepackage{lipsum}
\usepackage{multirow}
\usepackage{natbib}
\usepackage{paracol}
\usepackage{savesym}
\savesymbol{AND}

\theoremstyle{definition}
\newtheorem{theorem}{Theorem}
\theoremstyle{definition}
\newtheorem{lemma}[theorem]{Lemma}
\theoremstyle{definition}

\theoremstyle{definition}

\theoremstyle{definition}

\theoremstyle{definition}
\newtheorem{definition}[theorem]{Definition}
\theoremstyle{definition}
\newtheorem{proposition}[theorem]{Proposition}
\theoremstyle{definition}
\newtheorem{property}[theorem]{Property}
\theoremstyle{definition}
\newtheorem{example}[theorem]{Example}
\theoremstyle{definition}
\newtheorem{conjecture}[theorem]{Conjecture}
\theoremstyle{definition}
\newtheorem{remark}[theorem]{Remark}
\theoremstyle{definition}
\newtheorem{assumption}[theorem]{Assumption}
\theoremstyle{definition}
\newtheorem{corollary}[theorem]{Corollary}
\theoremstyle{definition}
\newtheorem{exercise}[theorem]{Exercise}
\theoremstyle{definition}
\newtheorem{simulation}[theorem]{Simulation}
\theoremstyle{definition}
\newtheorem{setting}{Setting}

 {\endMakeFramed}

\newenvironment{greenleftbar}{%
  \MakeFramed {\advance\hsize-\width \FrameRestore}}%
 {\endMakeFramed}

{\endMakeFramed}

\newenvironment{exercise-waku}
  {\begin{greenleftbar}\begin{exercise}}
  {\end{exercise}\end{greenleftbar}}
\newenvironment{simulation-waku}
  {\begin{greenleftbar}\begin{simulation}}
  {\end{simulation}\end{greenleftbar}}
\newenvironment{proposition-waku}
  {\begin{oframed}\begin{proposition}}
  {\end{proposition}\end{oframed}}
\newenvironment{definition-waku}
  {\begin{oframed}\begin{definition}}
  {\end{definition}\end{oframed}}
  \newenvironment{lemma-waku}
  {\begin{oframed}\begin{lemma}}
  {\end{lemma}\end{oframed}}
\newenvironment{remark-waku}
  {\begin{oframed}\begin{remark}}
  {\end{remark}\end{oframed}}
\newenvironment{theorem-waku}
  {\begin{oframed}\begin{theorem}}
  {\end{theorem}\end{oframed}}
\newenvironment{property-waku}
  {\begin{oframed}\begin{property}}
  {\end{property}\end{oframed}}
\newenvironment{corollary-waku}
  {\begin{oframed}\begin{corollary}}
  {\end{corollary}\end{oframed}}
  \newenvironment{example-waku}
  {\begin{oframed}\begin{example}}
  {\end{example}\end{oframed}}
  \newenvironment{conjecture-waku}
  {\begin{oframed}\begin{conjecture}}
  {\end{conjecture}\end{oframed}}
  \newenvironment{assumption-waku}
  {\begin{oframed}\begin{assumption}}
  {\end{assumption}\end{oframed}}
\newenvironment{setting-waku}
  {\begin{oframed}\begin{setting}}
  {\end{setting}\end{oframed}}

\newcommand{\x}{{\bm{x}}}

%% file: src/usonics2023style-005-frontmatter.tex
\begin{frontmatter}

  
  
  \title{Simulation-Aided Deep Learning for Laser Ultrasonic Visualization Testing}
  
  
  \author{Miya Nakajima\fnref{a1}\corref{cor1}}
  \author[a1]{Takahiro Saitoh}
  \author[a2]{Tsuyoshi Kato}
  
  \affiliation[a1]{organization={Graduate School of Science and Technology, Gunma University},
              addressline={Tenjin, Kiryu 376 8515, Japan}, 
              }
  \affiliation[a2]{organization={Faculty of Informatics, Gunma University},
              addressline={ Aramaki, Maebashi 371 8510, Japan}, 
              }
  \begin{abstract}
  In recent years, laser ultrasonic visualization testing (LUVT) has attracted much attention because of its ability to efficiently perform non-contact ultrasonic non-destructive testing.
  Despite many success reports of deep learning based image analysis for widespread areas, attempts to apply deep learning to defect detection in LUVT images face the difficulty of preparing a large dataset of LUVT images that is too expensive to scale. 
  To compensate for the scarcity of such training data, we propose a data augmentation method that generates artificial LUVT images by simulation and applies a style transfer to simulated LUVT images.
  The experimental results showed that the effectiveness of data augmentation based on the style-transformed simulated images improved the prediction performance of defects, rather than directly using the raw simulated images for data augmentation.
  \end{abstract}
  
  
  
  \begin{keyword}
  
  Laser ultrasonic visualization testing \sep Deep learning \sep Data augmentation \sep Style transfer
  
  \end{keyword}
\end{frontmatter}

%% file: src/usonics2023style-010-intro.tex
\section{Introduction}
In recent years, the demand for non-destructive testing for structural maintenance and management has increased due to the aging of structures and other societal factors.
In particular, ultrasonic non-destructive testing is superior in inspecting efficiency and safety and has been widely used. 
Among ultrasonic non-destructive testing methods, this paper focuses on the laser ultrasonic visualization testing (LUVT) technique~\cite{Kohler02usonics,Yashiro08ndtei}. 
LUVT can visualize the propagation of ultrasonic waves, and the presence or absence of defects can be determined from the image obtained by LUVT. 

However, in many ultrasonic non-destructive testing facilities, human inspectors still visually inspect for defects. 
In response to the ever-increasing demand for non-destructive testing, the field of non-destructive testing is plagued by a shortage of well-trained inspectors and an increased workload. 
As a solution to this issue, machine learning technology has recently been introduced to non-destructive ultrasonic testing to reduce the labor required for inspections~\cite{MENG2017,POSIC2022}. 

The goal of this study is to develop a machine learning technique for automatic inspection of images obtained by LUVT. 
The challenge of automating ultrasonic non-destructive testing in academia did not begin recently. It already has a long history and various learning machines such as shallow neural networks~\cite{Masnata96ndtei,Sambath10jnde,Ogi1990} and SVMs~\cite{Huifang20ndtei,xiaokai19ultras} have been examined.
In the early 2010s, rich deep models emerged and replaced standard methods in many applications such as image recognition, speech recognition, and natural language processing. 
Through the enrichment of machine learning models, the effectiveness of these tasks has been brought to human levels. 
These successes eventually motivated the researchers in ultrasonic non-destructive testing to try deep learning to automate inspections~\cite{CanteroChinchilla22ndtei,DaiQuocTran20sensors}.

Compared to generic image recognition tasks, however, the collection of training images for the non-destructive inspection is expensive and time-consuming, which limits the amount of available data for machine learning. 
However, the amount of data needed to train rich machine learning models such as deep neural networks is much larger than the amount needed to train human inspectors~\cite{Virkkunen2021virtual}. 
Hence, the scarcity of training data severely hampers the generalization performance of machine learning models that rely on large amounts of training data. 

This paper proposes an effective simulation-based data augmentation method to address the scarcity of training data for LUVT image inspection. 
The numerical simulations can generate the images for scattered wavefields from various defects that correspond to experimental results without having to cost LUVT measurements. 
However, the existence of differences between simulated and real images has a negative impact on machine learning, failing to build models that perform well for defect detection from unseen real images. 
The proposed method alleviates the differences by making simulated images more similar to real images by applying a style transfer technique~\cite{CycleGAN2017}. 
The experimental results reported in this paper shall demonstrate the effects of the simulation-based data augmentation combined with the style transfer. 

%% file: src/usonics2023style-210-da.tex
\begin{figure*}[t]
  \centering
  \begin{tabular}{lllll}
  (a) Original & (b) Flip & (b) Crop & (b) Grayscale & (b) ColorJitter
  \\
  \includegraphics[scale=0.58]{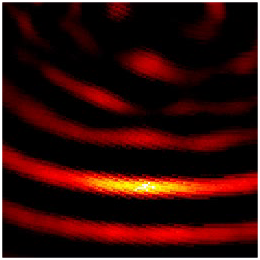}\hspace{3mm}
  &
  \includegraphics[scale=0.58]{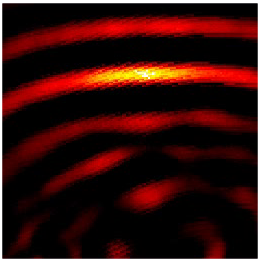}\hspace{3mm}
  &
  \includegraphics[scale=0.58]{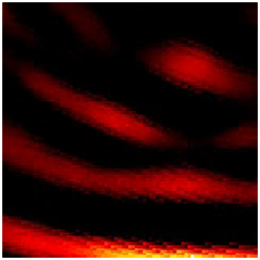}\hspace{3mm}
  &
  \includegraphics[scale=0.58]{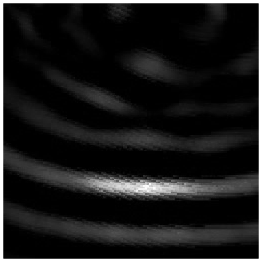}\hspace{3mm}
  &
  \includegraphics[scale=0.58]{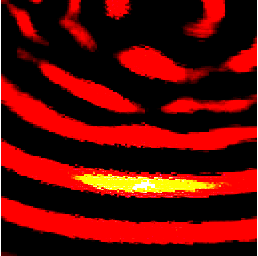}   
  \end{tabular}
  \caption{Data augmentation via simple transformation. \label{fig:data_augmentation}}
\end{figure*}

%% file: src/usonics2023style-020-related.tex
\section{Related Work}
Nowadays, few developers skip data augmentation to develop a generic image recognition algorithm.
This is because intuitive and natural approaches are easier to develop and implement for generic images than for other types of data. 
In general, data augmentation increases the diversity of the training images, thus reducing over-fitting to the training data and increasing generalization power.

\textbf{Simple Transformations: } 
Data augmentations commonly used for generic image recognition are simple image transformations, including geometric transformations such as flipping, rotating, cropping, cutout~\cite{devries2017cutout}, and intensity transformation such as hue and brightness changes. 
Examples of these typical data augmentations are shown in Figure~\ref{fig:data_augmentation}.
Karen et al.~\cite{brusimon2014very} demonstrated improved performance on the ImageNet dataset using these simple data augmentations. 

Another approach is to generate a new image by combining multiple images.
Mixup~\cite{zhang2018mixup} takes a linear combination of two images and category labels.  
CutMix~\cite{yun2019cutmix} improves Mixup and cutout by combining the two methods. 
Wang et al.~\cite{wang2021} verified the effect of using CutMix as a data augmentation in YOLOv4 on object detection performance. 

In addition, data augmentation methods using a generative adversarial network (GAN) ~\cite{goodfellow2014generative}, have also been proposed. 
GAN is a class of generative model that attempts to produce high-quality images via adversarial learning. 
Data augmentation using GANs has mainly been studied in the medical field using IAGAN~\cite{MOTAMED2021100779} and PGGAN~\cite{BowlChris2018}. 
However, data augmentation based on these image transformations produces images that could not possibly exist in reality, limiting the positive effect of classical data augmentation~\cite{DuWangzhe2019}. 

\textbf{Virtual Flaws: } To overcome the limitations of the simple transformation-based data augmentations, another approach called virtual flaws or virtual cracks were introduced for non-destructive inspection~\cite{Virkkunen2014ecndt}. 
The virtual flaw approach moves the flaw signal identified in real data to an arbitrary position to expand the training dataset. 
This approach has been used with success to train not only machine learning models but also human inspectors~\cite{Virkkunen2021virtual}. 
However, this approach is infeasible to LUVT images. It is because the change in an image due to a defect does not appear locally. Instead, its scattering wave spreads globally in the image, which makes it impossible to generate another defect image just by moving a local area to another position. 

\textbf{Simulation: } 
This study employs 
numerical simulations for data augmentation which possesses several advantages. 
The first advantage is the ability to generate images that meet the arbitrary intended purpose.
The simulator can generate a variety of images to meet specific conditions defined by the user.  
This is contrastive to data augmentation based on simple image transformations that are weak in synthesizing particular images specified by users. 
The second advantage is the ability to use the generated images on a large scale.
A simulator can acquire an unlimited number of images without incurring the cost of image acquisition.

Data augmentation by 
simulations has been proved to be effective in robot vision~\cite{anderson21crl}, automatic driving~\cite{Liu2020DataAT}, medical imaging~\cite{ALKL2023}, and defect detection on steel surfaces~\cite{Boikov21sym}. 
Their success motivated us to use simulation for data augmentation, although we found that the LUVT images generated by the 
numerical simulation have considerable differences from the actual measured images. This is because the actual LUVT images contain a lot of noise.
Due to the nature of these images, direct use of the corresponding LUVT images obtained by simulations
was not expected to result in good prediction performance.

\textbf{Style Transfer: }
In this study, to reduce the difference between simulator-generated images and real images, a style transfer technique is introduced. 
There have been proposed several style transformation algorithms~\cite{Gatys2016cvpr, ulyanov16icml, Elad2017tip, isola2017image, CycleGAN2017}. 
This study employs an unsupervised style transfer algorithm. 
Another type of style transfer algorithms is a supervised algorithm that learns with pairs of an input image and its style-transformed output image, and the correspondence is learned based on this information.
Isola et al.~\cite{isola2017image} developed Pix2Pix as a supervised style transfer and demonstrated its ability to transform grayscale images into color images.
However, in LUVT applications, it is difficult to prepare a pair of input images and their style-transformed output images in advance.
From this reason, a supervised style transfer was not employed in this study. 

%% file: src/usonics2023style-220-luvt.tex
\begin{figure}[t]
  \centering
\includegraphics[width=7cm]{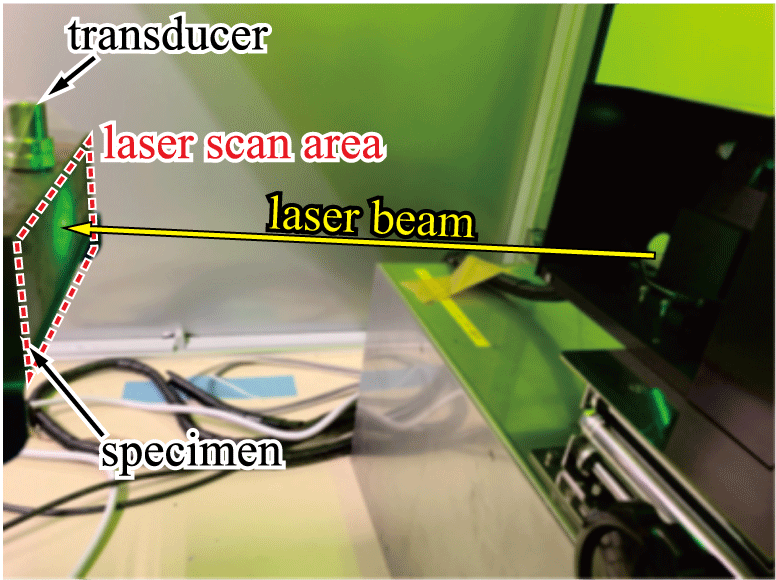}
\caption{LUVT device. By irradiating a laser beam on the surface of a specimen from the device on the right, the ultrasonic wave propagation on the irradiated surface can be imaged. 
\label{fig:luvt}}
\end{figure}

%% file: src/usonics2023style-270-alumspe.tex
\begin{figure}[t]
  \centering
  \includegraphics[width=7cm]{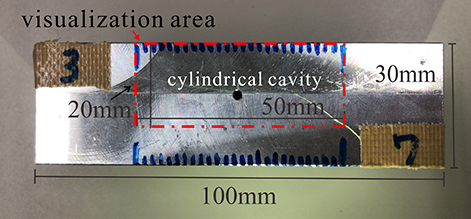}
  \caption{%
  Specimen. 
  \label{fig:alumspe}}
\end{figure}

%% file: src/usonics2023style-230-example.tex
\begin{figure}[t]
  \centering
  \begin{tabular}{ll}
  (a) Defect-free & (b) Defective
  \\
  \includegraphics[width=0.38\linewidth]{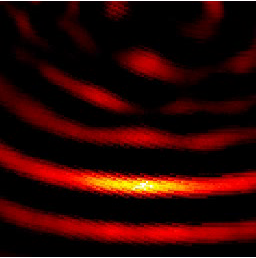}   
  &
  \includegraphics[width=0.38\linewidth]{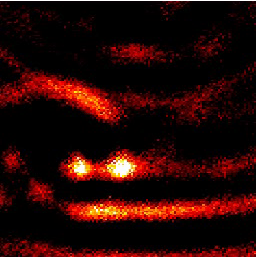}
  \\
  \end{tabular}
   \vspace*{-1mm}
  \caption{%
  Examples of real LUVT images. Panel (a) shows a defect-free image, in which ultrasonic waves incident from above propagate. Panel (b) shows an image with a defect, where scattered waves occurring at a defect can be seen in addition to the ultrasonic waves incident from above. \label{fig:example}}
\end{figure}

%% file: src/usonics2023style-030-dataset.tex
\section{LUVT and Preparing Dataset}
%
\textbf{Brief description of LUVT:} 
In non-destructive inspection using LUVT, a laser is first irradiated onto the test specimen to generate 
laser ultrasonic waves at the laser irradiation point, as shown in Figure~\ref{fig:luvt}. 
The excited laser ultrasonic waves are received by the pre-installed ultrasonic transducer.
This operation is performed for various laser irradiation points on the surface of a specimen.
After that, the reciprocity theorem~\cite{achenbach2003} is applied to the transducer and all irradiation points.
Then, an image as if ultrasonic waves were being transmitted can be obtained from the pre-installed ultrasonic probe~\cite{Yashiro08ndtei}.
This operation was applied to an aluminum specimen in Figure~\ref{fig:alumspe}, and the resulting image is shown in Figure~\ref{fig:example}.
For defect-free images, the incident ultrasonic waves from the probe placed above the specimen propagate as they are.
On the other hand, in the case of an image with a defect, scattered waves generated by the defect propagate isotropically in addition to the incident ultrasonic waves.
As a result, scattered waves from the defect are generated, as shown in Figure~\ref{fig:example}. The internal defects can be detected by identifying the existence of these scattered waves.

\textbf{Difficulty to Scale Dataset. }
Scattered waves from defects in images obtained by LUVT may not be visible even to humans due to measurement noise.
A large dataset for training is required to discriminate such images with high accuracy. 
A typical way of collecting a training dataset is to use the images obtained through the actual periodical inspection routine. 
A common obstacle of this approach is the scarcity of defect examples.
In this study, an alternative approach is adopted, which is to fabricate the defective sample in the following manner. 
\begin{enumerate}
  \item 
  Prepare a specimen.
  %
  \item 
  Artificially create defects inside the specimen using a drill. 
  \item 
  Obtain LUVT images using the imaging methods described above. 
\end{enumerate}
This procedure is repeated to obtain a positive example. 
To obtain a negative example, only steps 1 and 3 are performed.

In the experiments described in Section~\ref{s:exp}, aluminum specimens were used. 
A cylindrical cavity with a diameter of $d=2$mm was artificially created in each of the specimens.
The vertical, horizontal, and depth lengths for the aluminum specimen were 100mm, 30mm, and 50mm, respectively, as shown in Figure~\ref{fig:alumspe}. 
The longitudinal wave transducer with a center frequency of 2 MHz is used and located on the front side of the top surface, as found in the top area of Figure~\ref{fig:alumspe}. 
The red rectangular area ($20\text{mm}\times 50\text{mm}$) of this figure was the laser radiation area that corresponds to the LUVT visualization area for the ultrasonic wave propagation.  

Modern rich machine learning models cannot be powerful unless the training dataset can be scaled to a large size, although the specimens tend to be expensive. 
In addition, repeated measurements and imaging using the equipment require time and labor.
This is a barrier to scale LUVT data sets.

%% file: src/usonics2023style-240-artificial.tex
\begin{figure}[t]
    \centering
    \begin{tabular}{ll}
      (a) Defect-free  & (b) Defective
    \\
    \includegraphics[width=0.38\linewidth]{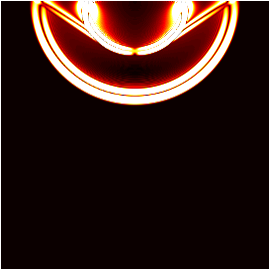}
       &
    \includegraphics[width=0.38\linewidth]{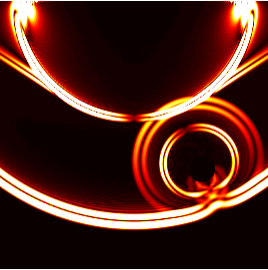}   
    \\
    \end{tabular}
    \caption{%
    Examples of simulated LUVT images. Panel (a) shows an image without defects and (b) shows an image with a defect. The difference in style from the real LUVT images is considerable. 
    \label{fig:artexample}}
  \end{figure}
  

%% file: src/usonics2023style-250-st.tex
\begin{figure}[t]
  \centering
      \includegraphics[scale=0.35]{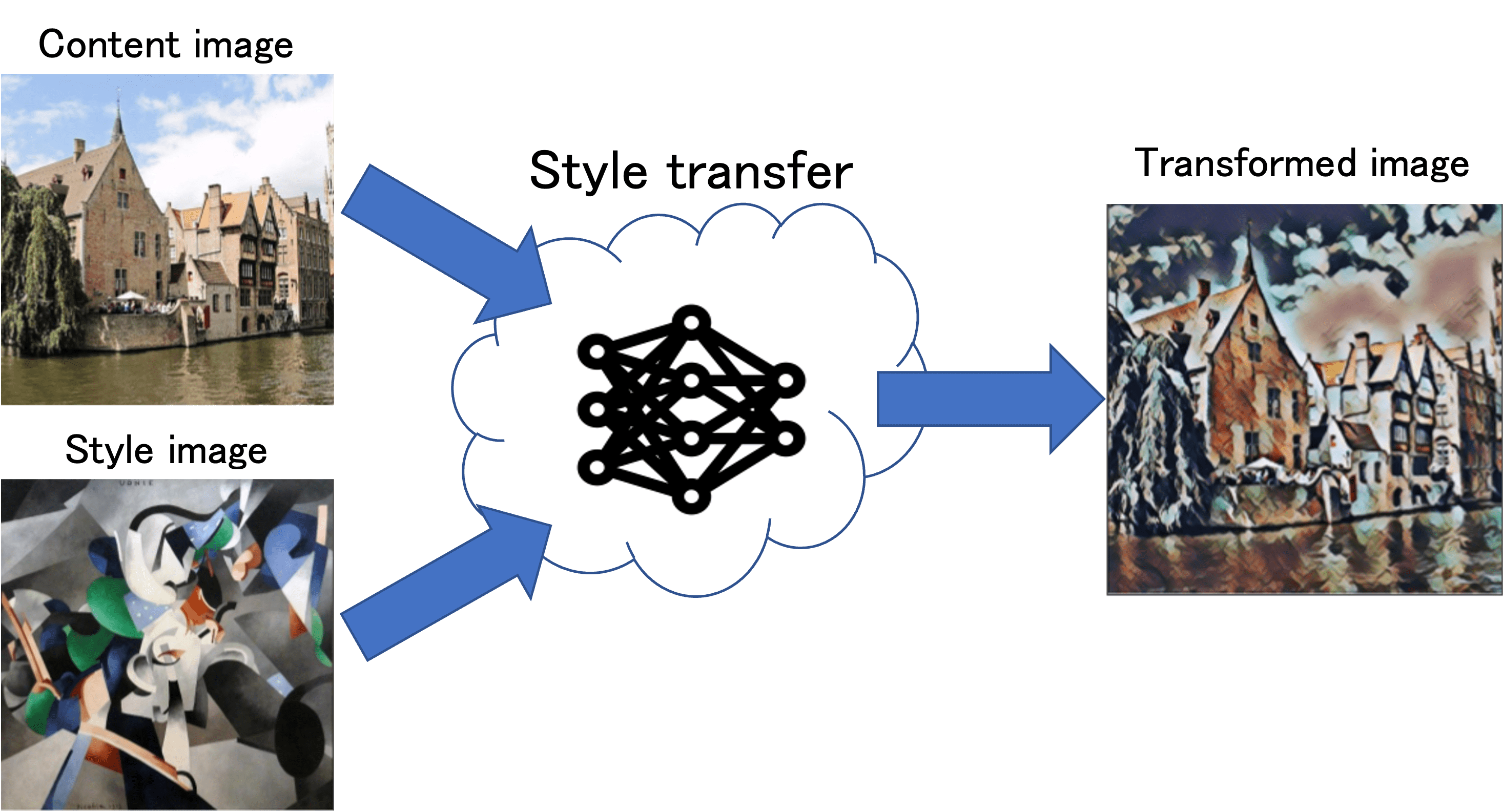}
      \caption{Example of style transformation. Source:\\ https://www.tensorflow.org/lite/examples/style\_transfer/overview.}
    \label{fig:st}
  \end{figure}
  

%% file: src/usonics2023style-260-prop.tex
\begin{figure}[t]
    \centering
    \includegraphics[scale=0.3]{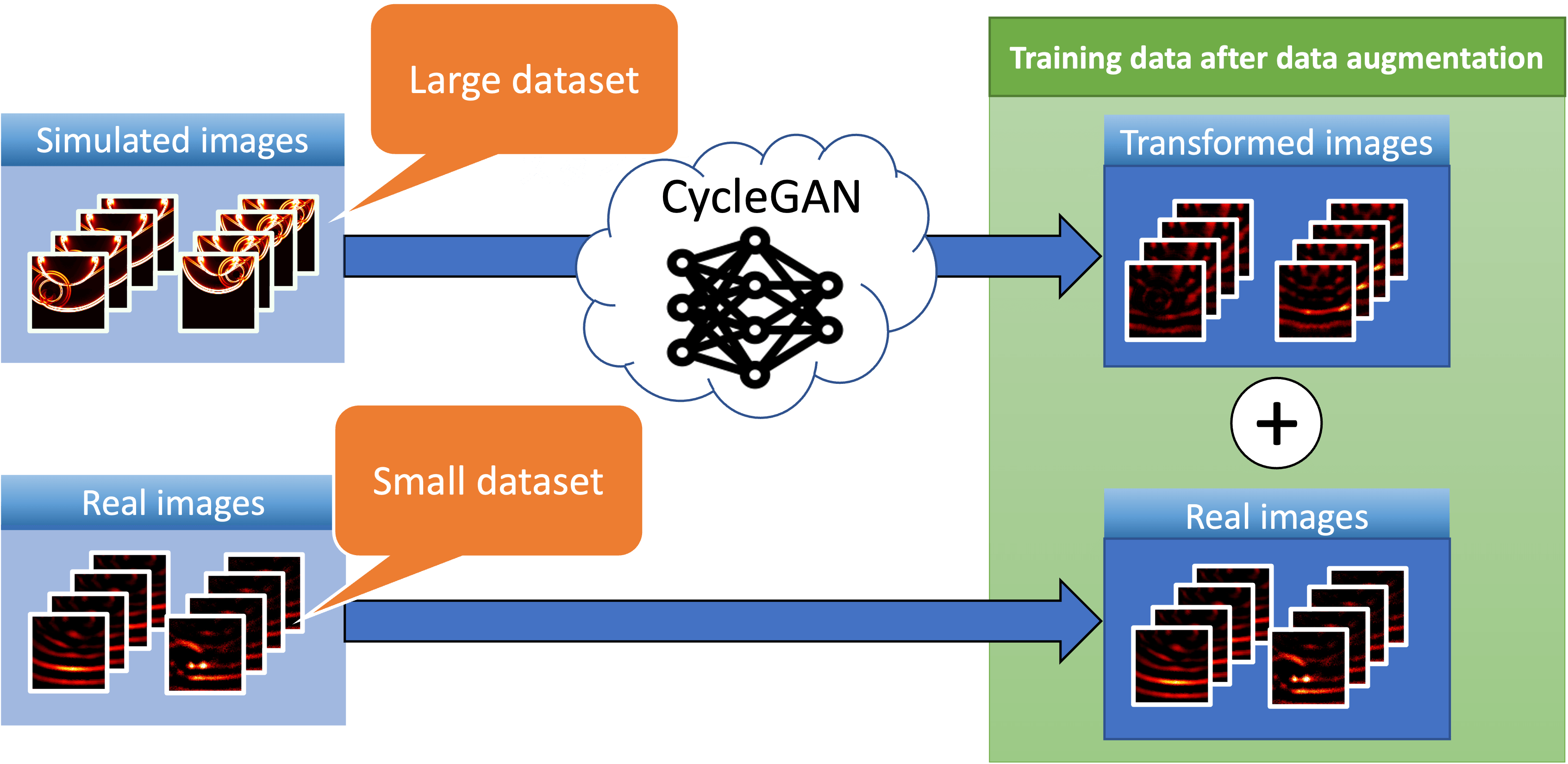}
    \caption{%
    Synthesized images from the physics simulation are refined using a style transfer to combine them with real images for training. 
    \label{fig:prop_method}}
    
  \end{figure}

%% file: src/usonics2023style-310-table-cygan.tex
\begin{table*}[t]
  \centering
  \caption{%
  Behaviors of a style transfer called CycleGAN. As the style transfer learned and reached 20,000 iterations, the simulated image approached the style of the real LUVT image. 
  \label{tab:cygan_result}}
  \scalebox{1}{
    \begin{tabular}{|c|cccc|}
      \hline Label & 0 iterations & 100 iterations & 10,000 iterations & 20,000 iterations \\ \hline
      \hline 
      Defect-free & 
      \begin{minipage}{30mm}
        \vspace{1mm}
        \centering
        \includegraphics[scale=0.5]{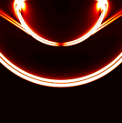}
        \vspace{1mm}
      \end{minipage} &
      \begin{minipage}{30mm}
        \centering
        \scalebox{0.5}{\includegraphics{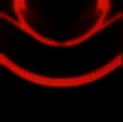}}
      \end{minipage} &
      \begin{minipage}{30mm}
        \centering
        \scalebox{0.5}{\includegraphics{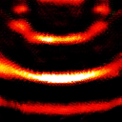}}
      \end{minipage} &
      \begin{minipage}{30mm}
        \centering
        \scalebox{0.5}{\includegraphics{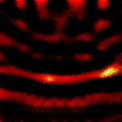}}
      \end{minipage} 
       \\ \hline
      Defective & 
      \begin{minipage}{30mm}
        \vspace{1mm}
        \centering
        \scalebox{0.5}{\includegraphics{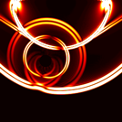}}
        \vspace{1mm}
      \end{minipage} &
      \begin{minipage}{30mm}
        \centering
        \scalebox{0.5}{\includegraphics{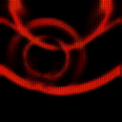}}
      \end{minipage} &
      \begin{minipage}{30mm}
        \centering
        \scalebox{0.5}{\includegraphics{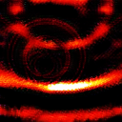}}
      \end{minipage} &
      \begin{minipage}{30mm}
        \centering
        \scalebox{0.5}{\includegraphics{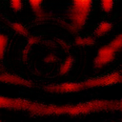}}
      \end{minipage} 
      \\ \hline
    \end{tabular}
    }
  \end{table*}

%% file: src/usonics2023style-040-meth.tex
\section{Proposed Data Augmentation Method}
%
%
\subsection{Image Generation From Simulation}
In order to use a physical simulator to synthesize an ultrasonic wave propagation image that is equivalent to an LUVT image, it is necessary to solve the elastic wave equation that is satisfied by the ultrasonic waves under the boundary conditions and initial conditions for the specimen.
Well-known methods used to solve the elastic wave equation include the finite difference method~\cite{HAGNESS2005}, the finite element method~\cite{hughes2012}, and the boundary element method~\cite{banerjee1981}.
The boundary element method is known as a high-precision wave analysis method, but it requires a relatively large amount of computation time.
On the other hand, the finite difference method and the finite element method can obtain numerical solutions in a relatively small computation time.

In this study, in order to prepare a large number of artificial images, simulation images were obtained using the time-domain finite difference method, which can obtain rough numerical solutions in a small computational cost.
The obtained simulated images are shown in Figure~\ref{fig:artexample}. 

\subsection{Style Transfer}
There is a difference between the LUVT image actually acquired and the image generated by the simulator.
Noise is introduced by various factors in the process of acquiring real LUVT images.
However, since the LUVT simulator is designed to understand physical phenomena, it cannot fully reproduce the process from ultrasonic scattering to image composition.
This difference causes domain shifts, and the image synthesized from the simulator does not improve generalization performance.
Therefore, rather than directly adding the simulator-generated images as training data, a transformation is applied to make it closer to the real image.

In this study, a technique called style transfer~\cite{CycleGAN2017}, which has been developed in the field of image recognition, is applied in order to make simulator-generated images more similar to real LUVT images.
Style transfer model learns a style to transform the style of input images (Figure~\ref{fig:st}).
In this paper, we propose introduction of a style transfer algorithm for the purpose of data augmentation of LUVT images. 
Style transfer model itself is needed to be trained in addition to the predictor of the existence of defect. 
The style transfer model learns the content domain from a set of content images and simultaneously learns the style domain from another image set.
Zhu et al.~\cite{CycleGAN2017} developed a style transfer called CycleGAN and demonstrated that the learning of these two domains can transform a painting style, such as a landscape photograph from an actual painting.
In the experiments reported in Section~\ref{s:exp}, we used CycleGAN to verify the effect of the proposed data augmentation method on predicting the defects from LUVT images.

%% file: src/usonics2023style-050-exp-cygan.tex
\section{Experiments}
\label{s:exp}
First, we illustrate how the style transfer changes the simulated images, and then we demonstrate the effect of the style transfer on defect detection performance. 

\subsection{Style Transfer}
In order to make the simulated images closer to the real images, a style transfer was performed using CycleGAN. 
To train CycleGAN, the simulated images were used for content images and the real LUVT images were used for style images.
The simulated LUVT dataset consists of 431 images for each of 55 defect locations.
The real LUVT dataset consists of 134 images for each of 203 defect locations. 
The optimization algorithm used to train CycleGAN is Adam with a learning rate of $0.0002$.
The number of iterations for CycleGAN training was $20,000$. 
The transformed image was finally obtained by the learning process.  

The images obtained from the style transfer of the simulated images using CycleGAN are shown in Table~\ref{tab:cygan_result}. 
The leftmost image is before applying the style transfer. 
The second, third, and fourth columns from the left show the transformed images after 100, 10,000, and 20,000 iterations of the learning process of CycleGAN, respectively.
Although each image shows differences in terms of wave intensities and defect visibility, it can be confirmed that the style of the LUVT image is getting closer to that of the LUVT image as CycleGAN is trained. 

In what follows, we shall demonstrate how data augmentation with style-transformed images affects the defect detection performance of the machine learning models.

%% file: src/usonics2023style-320-table-modeldetatil.tex
\begin{table*}[t]
  \centering
  \caption{Parameter set for training prediction models.~\label{tab:modeldetail}}
  \begin{tabular}{c||cccc}
    \hline
      &  EfficientNet & ResNeXt & ViT  \\
    \hline
    Input size & $(3,224,224)$ & $(3,224,224)$ & $(3,224,224)$  \\
    \# of epochs & 30 & 30 & 30   \\
    Batch size &  128& 128 & 256  \\
    Algorithm & Adam & Adam & AdamW \\
    Learning rate &  0.01 & 0.01 & 0.0001  \\
    \hline
  \end{tabular}
\end{table*}

%% file: src/usonics2023style-330-table-dataset.tex
\begin{table*}[t]
  \caption{LUVT Dataset~\label{table:dataset}}
  \centering
  \vspace{3mm}
  \begin{tabular}{c||cc}
    \hline
    & \# of defect-free images & \# of defective images \\
    \hline
    Real image  &  22,634 & 4,567    \\
    Simulated image  & 17,111 & 6,594  \\
    Transformed image & 17,111 & 6,594 \\
    \hline
  \end{tabular}
\end{table*}

%% file: src/usonics2023style-340-table-pred_result1.tex

\begin{table}[t]
  \caption{
  Defect detection performance when 61, 20, and 122 specimens are used for training, validation, and assessment, respectively. \label{tab:result1}}
  \begin{subtable}[t]{0.45\textwidth}
    \caption{EfficientNet\label{tab:effnet1}}
      \centering
      \scalebox{0.85}{
      \begin{tabular}{c|c|c|c|c}
       & Accuracy & Precision & Recall & F-score \\
      \hline \hline
      \textsc{Real alone} & 0.951 & 0.868 & 0.826 & 0.846\\
      \textsc{Real$+$Simulated} & 0.949 & 0.861 & 0.822 & 0.839\\
      \textsc{Proposed} & $\bm{0.961}$ & $\bm{0.899}$ & $\bm{0.859}$ & $\bm{0.878}$
     \end{tabular}
      }
  \end{subtable}
  \vspace{5mm}
  \vfill
  \begin{subtable}[t]{0.45\textwidth}
    \caption{ResNeXt\label{tab:ResNeXt1}}
    \centering
    \scalebox{0.85}{
      \begin{tabular}{c | c | c | c | c}
        & Accuracy & Precision & Recall & F-score \\
      \hline \hline
      \textsc{Real alone} & 0.954 & 0.873 & 0.842 & 0.857\\
      \textsc{Real$+$Simulated} & 0.950 & 0.849 & 0.849 & 0.848\\
      \textsc{Proposed} & $\bm{0.963}$ & $\bm{0.881}$ & $\bm{0.895}$ & $\bm{0.888}$
      \end{tabular}
    }
   \end{subtable}
   \vspace{5mm}
   \vfill
   \begin{subtable}[t]{0.45\textwidth}
    \caption{ViT\label{tab:ViT1}}
       \centering
       \scalebox{0.85}{
       \begin{tabular}{c|c|c|c|c}
         & Accuracy & Precision & Recall & F-score \\
       \hline \hline
       \textsc{Real alone} & 0.958 & 0.875 & 0.869 & 0.872 \\
       \textsc{Real$+$Simulated} & 0.961 & 0.887 & 0.876 & 0.882 \\
       \textsc{Proposed} & $\bm{0.968}$ & $\bm{0.900}$ & $\bm{0.903}$ & $\bm{0.901}$
       \end{tabular}
       }
    \end{subtable}
\end{table}

%% file: src/usonics2023style-350-table-pred_result2.tex

\begin{table}[t]
  \caption{
    Defect detection performance when 41, 20, and 142 specimens are used for training, validation, and assessment, respectively. \label{tab:result2}}
  \begin{subtable}[t]{0.45\textwidth}
    \caption{EfficientNet\label{tab:effnet2}}
      \centering
      \scalebox{0.85}{
      \begin{tabular}{c|c|c|c|c}
       & Accuracy & Precision & Recall & F-score \\
      \hline \hline
      \textsc{Real alone} & 0.935 & 0.810 & 0.789 & 0.799\\
      \textsc{Real$+$Simulated} & 0.937 & 0.813 & 0.806 & 0.808\\
      \textsc{Proposed} & $\bm{0.940}$ & $\bm{0.825}$ & $\bm{0.821}$ & $\bm{0.821}$
     \end{tabular}
     }
    \end{subtable}
    \vspace{5mm}
  \vfill
  \begin{subtable}[t]{0.45\textwidth}
    \caption{ResNeXt\label{tab:ResNeXt2}}
    \centering
    \scalebox{0.85}{
      \begin{tabular}{c | c | c | c | c}
        & Accuracy & Precision & Recall & F-score \\
      \hline \hline
      \textsc{Real alone} & 0.940 & 0.834 & 0.792 & 0.812\\
      \textsc{Real$+$Simulated} &  0.933 & 0.811 & 0.777 & 0.793\\
      \textsc{Proposed} & $\bm{0.949}$ & $\bm{0.844}$ & $\bm{0.847}$ & $\bm{0.845}$
      \end{tabular}
    }
   \end{subtable}
   \vspace{5mm}
   \vfill
   \begin{subtable}[t]{0.45\textwidth}
    \caption{ViT\label{tab:ViT2}}
       \centering
       \scalebox{0.85}{
       \begin{tabular}{c|c|c|c|c}
         & Accuracy & Precision & Recall & F-score \\
       \hline \hline
       \textsc{Real alone} & 0.945 & 0.858 & 0.800 & 0.827 \\
       \textsc{Real$+$Simulated} & 0.951 & 0.854 & 0.850 & 0.852 \\
       \textsc{Proposed} & $\bm{0.956}$ & $\bm{0.871}$ & $\bm{0.864}$ & $\bm{0.867}$
       \end{tabular}
       }
    \end{subtable}
\end{table}

%% file: src/usonics2023style-360-table-gradcam1.tex
\begin{table*}[!t]
  \caption{%
  Class activation mapping by Grad-CAM. Predictions for defect-free images are shown in the upper panel, and predictions for images with defects are shown in the lower panel.
  Only the proposed method correctly classified each image. \label{tab:gc1}}
  \centering
  \begin{tabular}{|c|cccc|}
    \hline  & Input & \textsc{Real alone} & \textsc{Real+Simulated} & \textsc{Proposed}\\ \hline
    \hline 
    Defect-free & 
    \begin{minipage}{30mm}
      \vspace{1mm}
      \centering
      \includegraphics[scale=0.6]{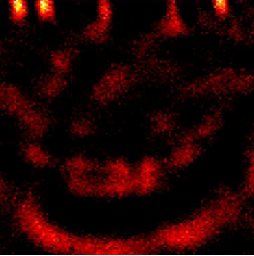}
      \vspace{1mm}
    \end{minipage} &
    \begin{minipage}{30mm}
      \centering
      \scalebox{0.6}{\includegraphics{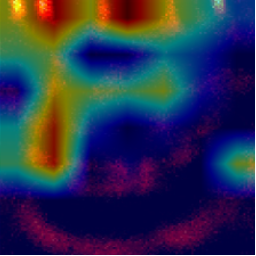}}
    \end{minipage} &
    \begin{minipage}{30mm}
      \centering
      \scalebox{0.6}{\includegraphics{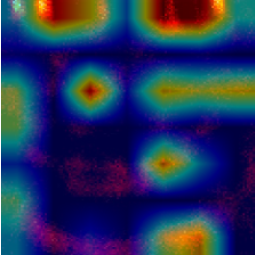}}
    \end{minipage} &
    \begin{minipage}{30mm}
      \centering
      \scalebox{0.6}{\includegraphics{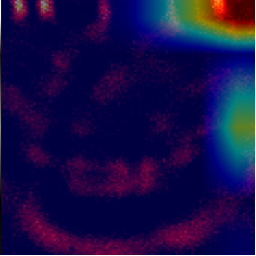}}
    \end{minipage} 
     \\ \hline
     Defective & 
    \begin{minipage}{30mm}
      \vspace{1mm}
      \centering
      \includegraphics[scale=0.6]{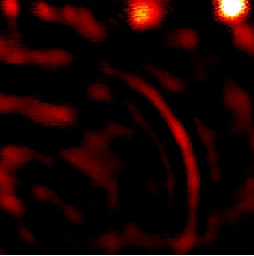}
      \vspace{1mm}
    \end{minipage} &
    \begin{minipage}{30mm}
      \centering
      \scalebox{0.6}{\includegraphics{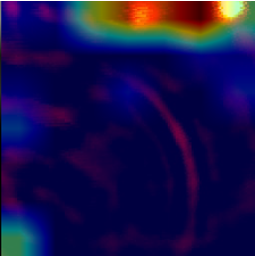}}
    \end{minipage} &
    \begin{minipage}{30mm}
      \centering
      \scalebox{0.6}{\includegraphics{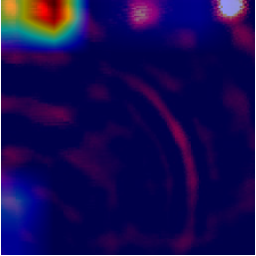}}
    \end{minipage} &
    \begin{minipage}{30mm}
      \centering
      \scalebox{0.6}{\includegraphics{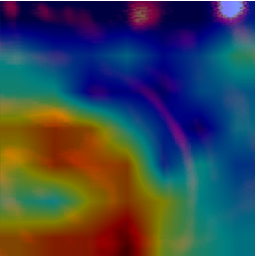}}
    \end{minipage} 
     \\ \hline
  \end{tabular}
\end{table*}

%% file: src/usonics2023style-060-exp-imgcls.tex
\subsection{Data Augmentation}
In order to demonstrate the effectiveness of data augmentation by style transfer, experiments were conducted comparing the case of LUVT images alone, data augmentation directly with simulated LUVT images, and data augmentation with style transformed images, denoted by \textsc{Real alone}, \textsc{Real$+$Simulated}, and \textsc{Proposed}, respectively. 
The performance of these three methods was verified using three prediction models including EfficientNet~\cite{pmlr-v97-tan19a}, ResNeXt~\cite{XieGDTH16}, and Vision Transformer (ViT)~\cite{dosovitskiy2021an}.
The hyper-parameters used in the training of each prediction model are shown in Table~\ref{tab:modeldetail}. 
In addition to simulation-based data augmentation, only HorizontalFlip was added for simple transformation-based data augmentation during training, with the weights pre-trained on ImageNet as initial values. 
The dataset used is summarized in Table~\ref{table:dataset}.
We pose binary categorization problem, where images up to the arrival of the wave to a defect are labeled as defect-free and images thereafter are defective. 

The performance evaluation was conducted as follows. 
First, real images from 203 specimens were divided into training data, validation data, and testing data.
For the method \textsc{Real alone}, the training data containing only the real images was used to train the prediction model. 
Method \textsc{Real+Simulated} mixed the simulated images to the training dataset, and \textsc{Proposed} method added the style-transformed images to the training dataset. 
For each epoch, the loss on the validation data was monitored, and the weight that minimized the loss was selected as the training result.
The weights obtained from the training were used to assess Accuracy, Precision, Recall, and F-score on the testing data. 
The above procedure was repeated five times with different random data partitioning patterns, and averaged over five data partitioning patterns to obtain the performance measures.

Table~\ref{tab:result1} reports the classification performance for the case where 203 image subsets were divided into 61 training subsets, 20 validation subsets, and 122 testing subsets. 
Let us first look at the data augmentation without style transfer. 
When using ViT for the prediction model, the \textsc{Real+Simulated} method achieved better performance than the \textsc{Real alone} method, although no improvement was observed when using EfficientNet and ResNeXt. 
This was because the simulated images were not similar enough to the real images to enhance the training of the prediction models. 
In contrast, the proposed method improved defect detection performance for all prediction models. 

Table~\ref{tab:result2} shows the prediction performances for the case where the number of training data is reduced by dividing the real LUVT image subsets into 41 training subsets, 20 validation subsets, and 142 testing subsets.  
It was observed from this table that even when the original training data set was smaller, mixing of real images and the transformed images can suppress the false detection and missed defects. 

\textbf{Visualization of decision making: }
All three prediction models we used have a deep structure. 
The internal workings of deep neural networks are not easily interpretable. 
A technique called Grad-CAM~\cite{Rampra2017} is a solution of this issue that provides a perspective for how decisions are made in deep networks.
Grad-CAM generates a heat map that represents the magnitude of the gradient obtained by back propagation.
The heat map is called a class activation map.
The deep prediction model is particularly concerned with areas of high gradient magnitude in its predictions.

The class activation mappings for the classification of LUVT images using the trained EfficientNet are shown in Table~\ref{tab:gc1}.
In the example of the prediction for defect-free images, the proposed method correctly determines the absence of defects. 
In contrast, with the other data augmentation methods, the prediction model looks at some areas in the image even though these areas are free of defects.
This tendency was especially observed when the ripples were disrupted independently of the defects.
Except for the proposed method, the approximate location of the defect was not identified, thereby leading to missing the defect for the input image depicted in Table~\ref{tab:gc1}.
Meanwhile, the prediction model trained with the proposed method gazed around the defect and successfully predicted the defect.
The proposed method thus provided better training data for the prediction models to more accurately determine the presence or absence of defects in LUVT images.

%% file: src/usonics2023style-070-concl.tex
\section{Conclusions}
In this paper, we proposed a method of data augmentation by applying a style transfer to simulated LUVT images. 
The collection of LUVT images is expensive and time-consuming, and therefore large training data is unavailable. 
One solution to this issue might be adding a variety of many simulation images to training data. 
The experiments conducted in this study showed that direct use of the simulated images was not effective because the simulated images greatly differ from the real LUVT images. 
To cope with this issue, we applied a style transfer to make the style of the simulated images closer to that of the LUVT images and then to add them to the training dataset.
The experiments demonstrated that the data augmentation based on the style-transformed simulated images improved the prediction performance of defects.

LUVT itself is still in its developmental stages, and improvements are being made.
Future work includes addressing more challenging machine learning problems that are expected to arise as measurement devices evolve.